\documentclass[espf]{aa}
\usepackage{graphicx}
\usepackage{amsmath}
\usepackage[psamsfonts]{euscript}
\usepackage[psamsfonts]{amssymb}
\usepackage{amsmath}
\usepackage{graphicx}
\usepackage{natbib}
\usepackage{epsfig}

\titlerunning{}

\def\kms{km s$^{-1}$}
\def\te{T_{\rm eff}}
\def\Mo{M_{\odot}}
\def\Lo{L_{\odot}}
\def\Ro{R_{\odot}}
\begin{document}
\title{High-resolution spectroscopy for Cepheids distance
determination}


\subtitle{II. A period-projection factor relation}
\titlerunning{A period-projection factor relation}
\authorrunning{N. Nardetto et al.}

\author{N. Nardetto \inst{1}, D. Mourard \inst{2}, Ph. Mathias \inst{2}, A. Fokin \inst{2,3}, D. Gillet
\inst{4}}

\institute{Max-Planck-Institut f\"{u}r Radioastronomie, Auf dem
H\"{u}gel 69, 53121 Bonn, Germany \and Observatoire de la C\^ote
d'Azur, Dpt. Gemini, UMR 6203, F-06130 Grasse, France \and Institute
of Astronomy of the Russian Academy of Sciences, 48 Pjatnitskaya
Str., Moscow 109017 Russia \and Observatoire de Haute Provence,
04870 Saint-Michel l'Observatoire, France }

\date{Received ... ; accepted ...}

\abstract {The projection factor is a key quantity for the
interferometric Baade-Wesselink (hereafter IBW) and
surface-brightness (hereafter SB) methods of determining the
distance of Cepheids. Indeed, it allows a consistent combination of
angular and linear diameters of the star.} {We aim to determine
consistent projection factors that include the dynamical structure
of the Cepheids' atmosphere.} {Hydrodynamical models of $\delta$~Cep
and $\ell$~Car have been used to validate a spectroscopic method of
determining the projection factor. This method, based on the
amplitude of the radial velocity curve, is applied to eight stars
observed with the HARPS spectrometer. The projection factor is
divided into three sub-concepts : (1) a geometrical effect, (2) the
velocity gradient within the atmosphere, and (3) the relative motion
of the ``optical'' pulsating photosphere compared to the
corresponding mass elements (hereafter $f_{\mathrm{o-g}}$). Both,
(1) and (3) are deduced from geometrical and hydrodynamical models,
respectively, while (2) is derived directly from observations.} {The
\ion{Fe}{I} 4896.439 \AA\, line is found to be the best one to use
in the context of IBW and SB methods. A coherent and consistent
period-projection factor relation (hereafter \emph{Pp} relation) is
derived for this specific spectral line: $p = [-0.064 \pm 0.020]
\log P + [1.376 \pm 0.023]$. This procedure is then extended to
derive dynamic projection factors for any spectral line of any
Cepheid.}{This \emph{Pp} relation is an important tool for removing
bias in the calibration of the period-luminosity relation of
Cepheids. Moreover, it reveals a new physical quantity
$f_{\mathrm{o-g}}$ to investigate in the near future.}


\maketitle

\section{Introduction}\label{s_Introduction}

The period-luminosity relation (hereafter \emph{PL} relation) of the
Cepheids is the basis of the extragalactic distance scale, but its
calibration is still uncertain at a $\Delta M = \pm 0.10$\,mag
level. Long-baseline interferometers currently provide a new,
quasi-geometric way to calibrate the Cepheids \emph{PL} relation.
Indeed, it is now possible to determine the distance of galactic
Cepheids up to $1$kpc with the interferometric Baade-Wesselink
method, hereafter IBW method; see for e.g. Sasselov \& Karovska
\cite{Sasselov94} and Kervella et al. \cite{Kervella04}.
Interferometric measurements lead to angular diameter estimations
over the whole pulsation period, while the stellar radius variations
can be deduced from the integration of the pulsation velocity. The
latter is linked to the observational velocity deduced from spectral
line profiles by the projection factor $p$. In this method, angular
and linear diameters have to correspond to the same physical layer
in the star to correctly estimate the distance. The projection
factor is currently the most important limiting quantity of the IBW
method. Indeed, in addition to limb-darkening effects, it is related
to the velocity gradient and, more generally, to the dynamical
structure of the Cepheid atmosphere.

In 1993, Butler studied the velocity gradient in the atmosphere of
four Cepheids using the excitation potential of spectral lines,
together with their asymmetry. Then, Butler et al. (1996) introduced
the velocity gradient in hydrostatic stellar atmosphere models and
found a 20\% reduction on the amplitude of the pulsation velocity
curve in the case of $\eta$~Aql. In addition, the $\gamma$-velocity
was reduced by 2\,\kms.

Fokin et al. (1996) studied the velocity gradient in the atmosphere
of $\delta$~Cep, based on hydrodynamical modelling. It was found to
be an important source of broadening for metallic spectral lines,
being similar to rotation or geometrical projection effects.

Using an improved version of the hydrodynamical model of $\delta$
Cep, Nardetto at al. (2004, hereafter Paper~I) proposed an
interferometric definition of the projection factor. A difference of
a few \kms was found between the amplitude of the photospheric and
line-forming region's velocity curves, leading to a bias of 5\% on
the derived distance. This theoretical result has been
observationally confirmed using trigonometric parallax measurements
of the HST and optical long baseline interferometry by M\'erand et
al. (2005).

Nardetto et al. \cite{Nardetto06a} show that spectro-interferometry
provides a new geometric view of the Cepheids' atmosphere. However,
the combination of different techniques (high-resolution
spectroscopy, spectro- and differential- interferometry) is needed
to efficiently constrain the physical parameters of the Cepheid
atmosphere and, in particular, the projection factor.

Recently, while comparing radial velocity curves of different
species, Petterson et al. (2005) found some evidence of a relation
between the velocity gradient in the Cepheids' atmosphere and their
period. Using a selected sample of absorption metallic lines, we
propose to probe the velocity gradient in the Cepheids' atmosphere
in order to determine {\it dynamic} projection factors. First, by
using $\delta$ Cep and $\ell$~Car hydrodynamic models, we present a
new spectroscopic method for determining the velocity gradient.
Then, the method is applied to the eight Cepheids observed with the
HARPS instrument. We discuss the choice of the spectral line and
then derive a specific period-projection factor relation (hereafter
\emph{Pp} relation). Results are discussed within the framework of
the \emph{PL} relation. Finally, we propose a general method (for
all lines and all stars) of determining the projection factor.

\section{Cepheids observed and selected spectral
lines.}\label{ss_Lines}

Ten stars have been observed with the HARPS spectrometer
($R=120000$) : \object{R~Tra}, \object{S~Cru}, \object{Y~Sgr},
\object{$\beta$~Dor}, \object{$\zeta$~Gem}, Y~Oph, \object{RZ~Vel},
\object{$\ell$~Car}, \object{RS~Pup}, and X~Sgr. In the first paper
of this series, Nardetto et al. (2006b, hereafter paper II) showed
that the radial velocity associated with the centroid of the
spectral line, together with the line asymmetry, are very important
tracers of the dynamical structure of the Cepheids' atmosphere.
X\,Sgr was studied separately by Mathias et al. (2006) because of
its very atypical behavior showing several components in the
spectral lines profiles. Y~Oph is not considered here due to its
insufficient phase coverage (see Paper\,II, Fig.\,3).

Using Kurucz's models \cite{kurucz92}, we identified about 150
unblended spectral lines. In Paper II, we considered only the
unblended metallic line \ion{Fe}{I} 6056.005\AA\,. In this second
paper, we have carefully selected 17 spectral lines following two
criteria: (1) in order to avoid bias in the determination of the
line depth, the continuum must be perfectly defined for all
pulsation phases and for all stars. An example of the quality
required is given in paper~II for the \ion{Fe}{I} 6056.005 \AA\
spectral line (see Fig.\,1); (2) the selected sample of lines has to
cover a wide range of depth. The selected spectral lines are
presented in Table\,\ref{Tab_Lines}. Depending on the star and the
spectral line considered, the line depth can range from 2\% to 55\%.

\begin{table}
\begin{center}
\caption{Spectral lines used in this study. \label{Tab_Lines}}
\begin{tabular}{lcccc}
\hline \hline \noalign{\smallskip}

Name    &   Wavelength (\AA)       \\
    &           \\
    \hline
\ion{Fe}{I} &   4683.560     \\
\ion{Fe}{I} &   4896.439      \\
\ion{Fe}{I} &   5054.643      \\
\ion{Ni}{I} &   5082.339    \\
\ion{Fe}{I} &   5367.467     \\
\ion{Fe}{I} &   5373.709      \\
\ion{Fe}{I} &   5383.369     \\
\ion{Ti}{II}&   5418.751      \\
\ion{Fe}{I} &   5576.089      \\
\ion{Fe}{I} &   5862.353     \\
\ion{Fe}{I} &   6024.058      \\
\ion{Fe}{I} &   6027.051     \\
\ion{Fe}{I} &   6056.005      \\
\ion{Si}{I} &   6155.134     \\
\ion{Fe}{I} &   6252.555     \\
\ion{Fe}{I} &   6265.134      \\
\ion{Fe}{I} &   6336.824     \\

\hline \noalign{\smallskip}
\end{tabular}
\end{center}
\end{table}

\section{Hydrodynamical models.} \label{ss_modeling}

The hydrodynamical model of $\delta$~Cep is presented in Paper~I. In
addition, we have derived a new, consistent model of $\ell$~Car.
Since the main stellar quantities of $l$\,Car (HD\,84810) are still
uncertain, we tried several sets of luminosity $L$, effective
temperature $\te$, and mass $M$ in order to get suitable
observational quantities, such as the pulsation period $P$, the
average radius of the star $\overline{R}$, bolometric and radial
velocity curves, and the line profiles. The \emph{ML} relation was
taken from Chiosi et al. (1993), and the OPAL opacity tables (Rogers
\& Iglesias (1992)) were used.

This leads to the following set of parameters for a 154-zone model:
$M=11.5\,\Mo$, $L=21000\,\Lo$, $\te=5225 K$, $Y=0.28$, and $Z=0.02$,
which corresponds to a typical Pop. I chemical composition. The
inner boundary has been fixed at about $T=3.5\cdot 10^6$\,K,
corresponding to 4\,\% of the photospheric radius, so the model
envelope with the atmosphere contains about 53\,\% of the stellar
mass. The atmosphere itself contains about $10^{-4}$ of the total
stellar mass.

We started the hydrodynamical calculations with a linear,
non-adiabatic fundamental-mode velocity profile having a value of 10
\,\kms\, at the surface. At the limit cycle, the pulsation period is
$34.4$~days, very close (3\%) to the observational ($P=35.6$~days)
value deduced by Szabados et al. \cite{Szabados89}. Bolometric and
radial velocity amplitudes are $1.3$ mag and $50$\,\kms,
respectively. The relative radius amplitude at the surface is
$\Delta R/R=17\,\%$. The mean photospheric radius is about
$\overline{R}=180\,\Ro$.

For the hydrodynamical models of $\delta$~Cep and $\ell$~Car used in
this paper (see Sect. \ref{s_Method}), a careful analysis of the
dynamical structure of their atmospheres was performed based on
radiative transfer computations (under local thermal equilibrium) of
all spectral lines in Table\,\ref{Tab_Lines}.

\begin{table}
\begin{center}
\caption[]{Hydrodynamical models of Cepheids.

\label{Tab_Cepheids_Parameters}}
\begin{tabular}{lccccc}
\hline \hline \noalign{\smallskip}

Name     &  $P$     &   $T_{eff}$  & $\frac{L}{L_{\odot}}$              &        $\frac{M}{M_{\odot}}$        &         $\frac{\overline{R}}{R_{\odot}}$                        \\
         &     [days]   &     [K]          &                              &                                   &                                                      \\
\hline

S Cru       & 4.7  &   5900     &  1900         &       5.6               &               42  \\
$\delta$~Cep & 5.4 &  5877   &    1995 &        4.8 &     43 \\
Y Sgr        & 5.7   &   5850     &  2200         &       5.0               &            45    \\
$\beta$~Dor & 9.9   &   5500    &  3500         & 5.5                     & 65                \\
$\zeta$~Gem & 10.4  &   5500    &  3600         & 5.0                      & 64               \\
RZ Vel      & 21.6   &   5400    &  7450         & 7.0                     &  109              \\
$\ell$~Car      & 34.4   &   5225    &  21000    & 11.5                    &  180              \\
RS Pup      & 42.9   &   5100    &  22700         & 9.7                     &  186              \\
\hline \noalign{\smallskip}
\end{tabular}
\end{center}
\end{table}

We only derived the photospheric pulsation velocity for other
Cepheids (see Sect. \ref{ss_fog}). Thus, the physical parameters are
only roughly estimated since the dynamical structure of the
Cepheids' atmosphere is not considered. The period and the radius of
the Cepheids are consistent with observations at the 2\% and 3\%
levels respectively, while the mass follows the period-mass relation
of Choisi et al. (1993) within 13\%. The effective temperature and
luminosity were mostly constrained by the period and the radius.
R~TrA was not modeled because of its extremely short period. A
specific and in-depth study would be necessary to model this star.

The hydrodynamical models are presented in Table
\ref{Tab_Cepheids_Parameters}.

\section{New insights into understanding the projection factor} \label{s_Method}

In this section, we propose a division of the projection factor into
sub-concepts in order to allow a direct constraint from HARPS
spectroscopic observations. To test this method, we consider the
hydrodynamic models of $\delta$~Cep and $\ell$~Car.

\subsection{The projection
factor definition}\label{sss_pf_ref}

First of all, we have to provide a definition for the projection
factor that should be applied in the IBW and SB methods. This has
already be done in Paper I, but it must now be refined and adapted
to the method proposed here.

We define the interferometric projection factor as
\begin{equation} \label{Eq_pf}
p=\frac{\Delta V_{\mathrm{p}}^{\mathrm{o}}}{\Delta RV_\mathrm{c}}
\end{equation} where $\Delta V_{\mathrm{p}}^{\mathrm{o}}$ is the amplitude of the
pulsation velocity curve associated to the photosphere (subscript
$p$) of the star. It corresponds exactly to the {\it optical}
(subscript $o$) barycenter of the photosphere defined by
$\tau_{\mathrm{c}}=2/3$, where $\tau_\mathrm{c}$ is the optical
depth in the continuum. $\Delta RV_{\mathrm{c}}$ is the amplitude of
the radial velocity curve obtained with the centroid method, i.e.
the first moment of the spectral line. This definition is justified
for the following reasons:

First, we consider the pulsation velocity instead of the radius, as
already proposed in paper\,I, in order to allow a direct application
to spectroscopic observations.

Second, we consider velocity amplitudes to avoid difficulties
related to the $\gamma$-velocity. The $\gamma$-velocity is the
averaged value of the radial velocity curve over one pulsation
period. This quantity depends on the line considered. Moreover, as
shown in Paper\,I, the projection factor is mainly constrained by
velocity amplitudes, which also justifies this choice.

Third, we consider the $RV_{\mathrm{c}}$ velocity instead of the
radial velocity associated to the Gaussian fit method, as in Paper
I. This is required for obtaining a projection factor independent of
the rotation of the star and the natural width of the spectral lines
(see Fig.~8, Paper~II). We insist on this definition of the radial
velocity since it is absolutely required to allow important
comparisons between the projection factors of Cepheids.

In the next section, we consider the \ion{Fe}{I} 4896.439 \AA\
spectral line as a reference. But the results, in terms of
consistency, can be generalized to any other spectral line. We find
$p=1.33$ for $\delta$~Cep  and $p=1.27$ for $\ell$~Car. These values
are our references in the following. If we apply a minimization
process between $V_{\mathrm{p}}^{\mathrm{o}} (\phi)$ and
$pRV_{\mathrm{c}} (\phi)$, with $p$ the only free parameter, we find
differences of $0.01$, which provides a good estimate of the
uncertainty on the projection factor.

\subsection{Decomposition of the projection
factor}\label{sss_pf_decomposition}

We now divide the projection factor:

\begin{equation} \label{Eq_pf_decomposition}
p= p_{\mathrm{o}}\,f_{\mathrm{grad}}\,f_{\mathrm{o-g}}
\end{equation}

into different quantities where,

\begin{itemize}

\item $f_{\mathrm{o-g}}=\frac{\Delta V_{\mathrm{p}}^{\mathrm{o}}}{\Delta
V_{\mathrm{p}}^{\mathrm{g}}}$, where $\Delta
V_{\mathrm{p}}^{\mathrm{g}}$ is the gas (subscript $g$) velocity
associated to the {\it optical} barycenter ($\tau_{\mathrm{c}}=2/3$)
of the photosphere. Thus, $f_{\mathrm{o-g}}$ is linked to the
distinction between the {\it optical} and {\it gas} photospheric
layers. The {\it optical} layer is the location where the continuum
and line photons are generated (e.g. the location of the
photosphere). The {\it gas} layer is the location of some mass
element in the hydrodynamic model mesh where, at some moment in
time, the photosphere is located. Given that the location of the
photosphere moves through different mass elements as the star
pulsates, the two ``layers'' have different velocities, hence the
necessity of the $f_{\mathrm{o-g}}$ definition. Indeed, the
interferometer in the continuum is only sensitive to the {\it
optical} layer.

\item $f_{\mathrm{grad}}= \frac{\Delta V_{\mathrm{p}}^{\mathrm{g}}}{\Delta V_{\mathrm{l}}^{\mathrm{g}}}$, where
$\Delta V_{\mathrm{l}}^{\mathrm{g}}$ is the {\it gas} velocity
associated to the optical barycenter ($\tau_{\mathrm{l}}=2/3$) of
the line-forming (subscript~$l$) region. Thus, $f_{\mathrm{grad}}$
is linked to the velocity gradient in the atmosphere of the star.
This quantity depends on the line considered.

\item $p_{\mathrm{o}}= \frac{\Delta V_{\mathrm{l}}^{\mathrm{g}}}{\Delta RV_{\mathrm{c}}}$
is the geometrical projection factor. It corresponds to an
integration of the pulsation velocity field (associated with the
line-forming region) projected on the line of sight and weighted by
the surface brightness of the star (including limb-darkening in the
spectral line). To derive $p_{\mathrm{o}}$, we use intensity
distributions in the continuum provided by the model. As described
in Nardetto et al. \cite{Nardetto06a}, there is a relation between
the pulsation velocity and the limb-darkening (and thus
$p_{\mathrm{o}}$): the intensity profile corresponding to the
highest velocity {\it at contraction} is the most limb-darkened,
while the profile corresponding to the highest velocity {\it at
expansion} is the least limb-darkened. Given that our definition of
$p_{\mathrm{o}}$ is related to the amplitude of radial and pulsation
velocity curves, we consider the median value (peak-to-peak) of the
$p_{\mathrm{o}}(\phi)$ curve to derive the $p_{\mathrm{o}}$-factor.
However, considering intensity distribution {\it in the continuum}
is an approximation. Indeed, hydrodynamic effects can result in much
limb-darkening, especially at the wavelengths corresponding to
spectral lines (see e.g. Marengo et al. 2003). Nevertheless, this
seems to be negligible since we obtain a good decomposition of the
projection factor.

\end{itemize}

To test this decomposition, we deduced these three quantities
directly from the hydrodynamical model, considering the \ion{Fe}{I}
4896.439 \AA\ metallic line. For $\delta$~Cep, we find
$f_{\mathrm{o-g}}~=0.963$, $f_{\mathrm{grad}}~=0.993$, and
$p_{\mathrm{o}}=1.390$; and for $\ell$~Car
$f_{\mathrm{o-g}}~=0.944$, $f_{\mathrm{grad}}~=0.982$, and
$p_{\mathrm{o}}=1.366$. The corresponding projection factors are
$p$[$\delta$~Cep]$=0.963*0.993*1.390=1.33$ and
$p$[$\ell$~Car]$=0.944*0.982*1.366=1.27$, which correspond to our
reference values. The physical decomposition of the projection
factor is thus consistent.

\begin{figure*}[]
\resizebox{\hsize}{!}{\includegraphics[clip=true]{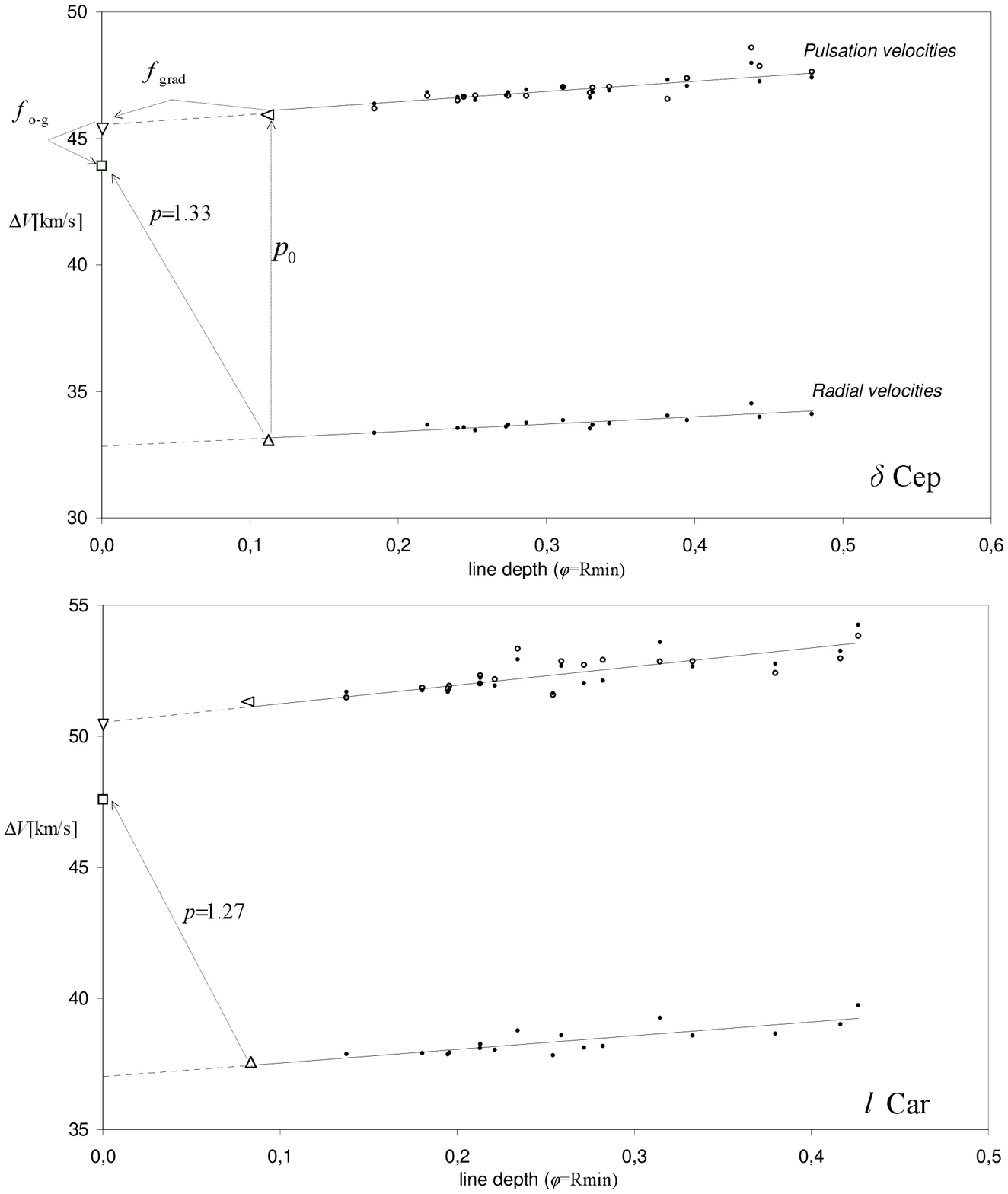}}
\caption{The projection factor decomposition ($p=p_{\mathrm{o}}
f_{\mathrm{grad}} f_{\mathrm{o-g}}$) in the case of the
\ion{Fe}{I}~4896.439 \AA\ spectral line: $\Delta
RV_{\mathrm{c}}$~({\tiny $\triangle$}), $\Delta
V_{\mathrm{l}}^{\mathrm{g}}$~($\triangleleft$), $\Delta
V_{\mathrm{p}}^{\mathrm{g}}$~($\triangledown$) and $\Delta
V_{\mathrm{p}}^{\mathrm{o}}$~({\tiny $\Box$}). Reference values
$p=1.33$ and $p=1.27$ are indicated for $\delta$~Cep and $\ell$~Car,
respectively. The proposed method of the $f_{\mathrm{grad}}$
determination is to first derive $\Delta RV_{\mathrm{c}}$ as a
function of the line depth for all spectral lines (lower part, black
points). Then, when translating ($D$, $\Delta RV_{\mathrm{c}}$)
points into ($D$, $p_{\mathrm{o}} \Delta RV_{\mathrm{c}}$), a new
linear relation is found (upper part of the figure, black points).
This new relation is (1) coherent with the pulsation velocity
gradient in the atmosphere (open circles) and (2) its zero-point is
consistent with the pulsation velocity corresponding to the {\it
gaseous} layer of the photosphere: $\Delta
V_{\mathrm{p}}^{\mathrm{g}}$~($\triangledown$). } \label{Fig_method}
\end{figure*}

One can notice that the $p$-factors presented in this paper are
independent of the pulsation phase. Indeed, they correspond to a
specific definition based on velocity amplitudes (Eq. \ref{Eq_pf}).
Figure\,\ref{Fig_method} represents the projection factor
decomposition in the case of the \ion{Fe}{I}~4896.439 \AA\ line for
$\delta$~Cep and $\ell$~Car: $\Delta RV_{\mathrm{c}}$~({\tiny
$\triangle$}), $\Delta
V_{\mathrm{l}}^{\mathrm{g}}$~($\triangleleft$), $\Delta
V_{\mathrm{p}}^{\mathrm{g}}$~($\triangledown$), and $\Delta
V_{\mathrm{p}}^{\mathrm{o}}$~({\tiny $\Box$}).

\subsection{Is there a consistent way to derive
$f_{\rm grad}$ directly from observations?}\label{sss_method}

Let assume a linear relation between the line depth and the position
of the line-forming region in the atmosphere of the star. Then,
considering different spectral lines spread all over the atmosphere,
it should be possible to determine the velocity gradient within the
atmosphere. Moreover, by extrapolation to the zero line depth, it
should be possible to reach the photospheric pulsation velocity.

To test these ideas, we consider the spectral line depth
corresponding to the minimum extension of the star ($D$ in the
following). We discuss this choice below. Then, we derive $\Delta
RV_{\mathrm{c}}$ for all spectral lines in Table \ref{Tab_Lines}.
From this quantity, one should be able to probe the velocity
gradient in the atmosphere and {\it directly} derive
$f_{\mathrm{grad}}$. For that, we plot $\Delta RV_{\mathrm{c}}$ as a
function of the line depth for all spectral lines
(Fig.~\ref{Fig_method}). We find a linear correlation ($\Delta
RV_{\mathrm{c}} = a_0 D + b_0$) given by the following relations~:
$\Delta RV_{\mathrm{c}}$[$\delta$~Cep]$ = 2.90D + 32.84$ and $\Delta
RV_{\mathrm{c}}$[$\ell$~Car]$ = 5.20D + 37.02$. Interestingly, these
relations are spectroscopic observables.

Then, following our decomposition, we translate ($D$, $\Delta
RV_{\mathrm{c}}$) points into ($D$, $p_{\mathrm{o}} \Delta
RV_{\mathrm{c}}$) with $p_{\mathrm{o}}$[$\delta$~Cep]$=1.390$ and
$p_{\mathrm{o}}$[$\ell$~Car]$=1.366$ (Fig.~\ref{Fig_method}). The
new relations are then : $p_{\mathrm{o}} {\Delta
RV_{\mathrm{c}}}$[$\delta$~Cep] $ = 4.03D + 45.64$ and
$p_{\mathrm{o}} {\Delta RV_{\mathrm{c}}}$[$\ell$~Car]$ = 7.10D +
50.53$. Another possible way to derive $p_{\mathrm{o}}$, instead of
using hydrodynamical modeling, is to consider a linear law for the
continuum-intensity distribution of the star defined by
$I(\cos(\theta))=1-u_{\mathrm V}+u_{\mathrm V}\cos(\theta)$, where
$u_{\mathrm V}$ is the limb-darkening of the star in the V band
(Claret et al. 2000), and $u_{\mathrm V}$ is related to the
effective temperature $T_{\mathrm{eff}}$, the surface gravity $\log
g$, the star metallicity, and the turbulent velocity. Using this
method to derive $p_{\mathrm{o}}$, we find that a slight correction
must be applied to allow a comparison between geometric and
hydrodynamic modeling. We find
$p_{\mathrm{o}}$[hydro]$=p_{\mathrm{o}}$[geo]$-0.010$ for
$\delta$~Cep and
$p_{\mathrm{o}}$[hydro]$=p_{\mathrm{o}}$[geo]$-0.025$ for
$\ell$~Car. As already mentioned, the limb-darkening in the
continuum used to derive $p_{\mathrm{o}}$[hydro] is linked to the
pulsation velocity and, more generally, to the whole dynamical
structure of the Cepheid atmosphere. Such a dynamical effect could
explain the difference that we find between $p_{\mathrm{o}}$[hydro]
and $p_{\mathrm{o}}$[geo]. From this, we estimate a $0.01$
uncertainty on $p_{\mathrm{o}}$. The limb-darkening (and thus
$p_{\mathrm{o}}$) will be studied in detail by interferometry in the
near future.

We then overplot the amplitude of the modeled pulsation velocity
curves corresponding to the different line-forming regions $\Delta
V_{\mathrm{l}}^{\mathrm{g}}$. Two important points have to be
mentioned here.

\begin{figure*}[]
\resizebox{\hsize}{!}{\includegraphics[clip=true]{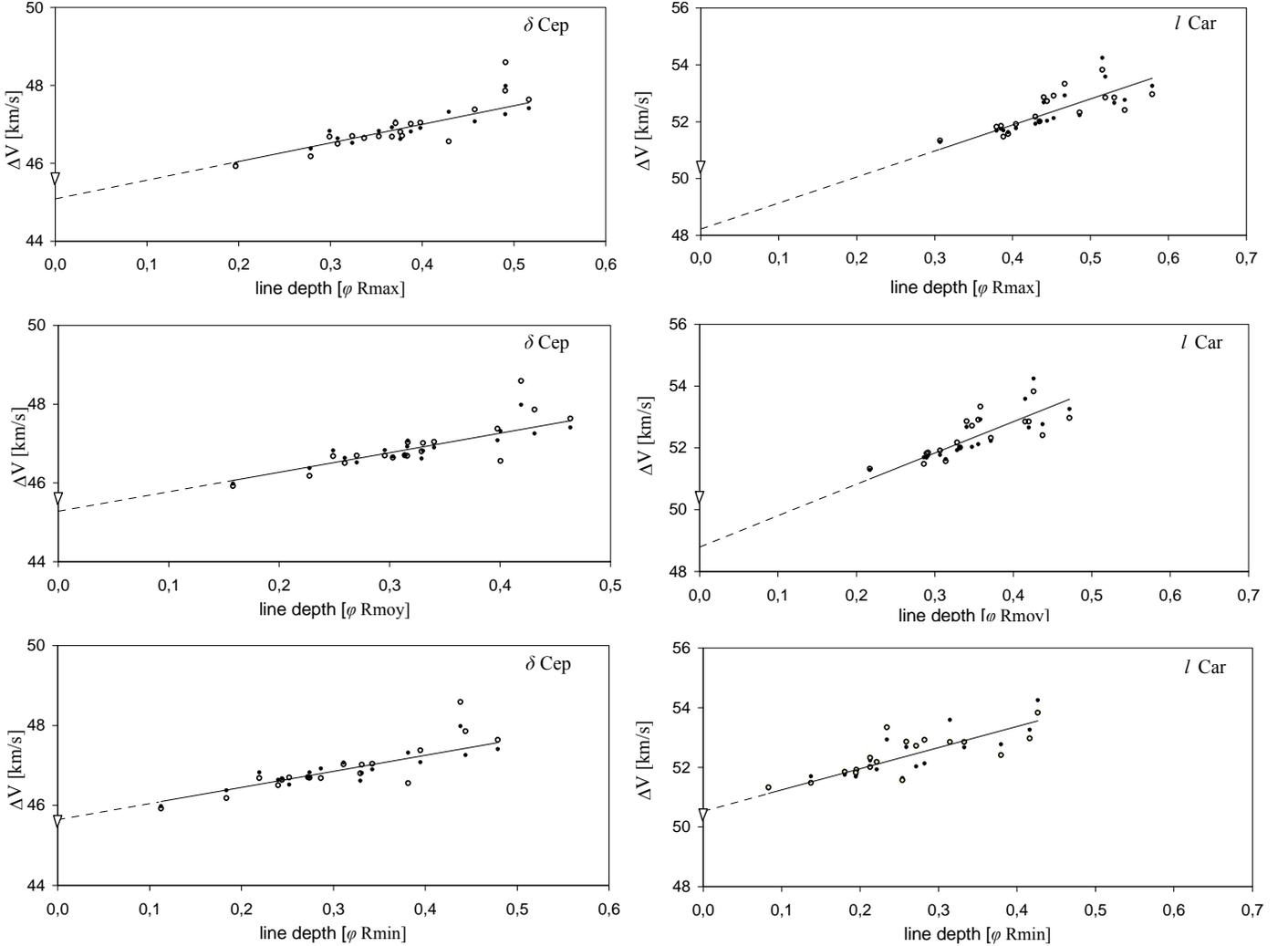}}
\caption{$\Delta V_{\mathrm{l}}^{\mathrm{g}}$ (open circles) as a
function of the line depth estimators D($\phi=R_{\mathrm{max}}$),
$<D>$, and D($\phi=R_{\mathrm{min}}$). Black points correspond to
the $p_{\mathrm{o}} \Delta RV_{\mathrm{c}}$ quantity. The symbol
$\triangledown$ indicates $\Delta V_{\mathrm{p}}^{\mathrm{g}}$;
i.e., the amplitude of the photospheric pulsation velocity
associated to the {\it gas}. We find good agreement between $\Delta
V_{\mathrm{l}}^{\mathrm{g}}$ and $p_{\mathrm{o}} \Delta
RV_{\mathrm{c}}$, which both present an interesting linear relation.
Through extrapolation (dashed line), we find that the best estimator
of the line-forming region is the line depth associated with the
minimum radius, a result confirmed for $\delta$~Cep and $\ell$~Car
hydrodynamic models.} \label{Fig_depth}
\end{figure*}

\begin{enumerate}
\item The superimposition of $p_{\mathrm{o}} \Delta RV_{\mathrm{c}}$ and $\Delta
V_{\mathrm{l}}^{\mathrm{g}}$ values is very satisfactory for both
stars. Thus, even if the $\Delta RV_{\mathrm{c}}$ quantity includes
all the dynamical structure of the line-forming region (including
the limb darkening), it seems, on average, to give direct access to
the {\it gas} velocity corresponding to $\tau_{\mathrm{l}}=2/3$,
i.e. $\Delta V_{\mathrm{l}}^{\mathrm{g}}$. In particular, the limb
darkening (and thus the $p_{\mathrm{o}}$-factor) seems to be
independent of the spectral line considered. For any line depth $D$,
we still have $\Delta V_{\mathrm{l}}^{\mathrm{g}} = p_{\mathrm{o}}
\Delta RV_{\mathrm{c}}$.

\item  We find that the zero-points of the ($D$, $p_{\mathrm{o}} \Delta RV_{\mathrm{c}}$)
relations, $45.64$ [$\delta$~Cep] and $50.53$ [$\ell$~Car], are
close to the reference values derived directly from the pulsation
velocities of the model, $\Delta
V_{\mathrm{p}}^{\mathrm{g}}=45.60$[$\delta$~Cep] and
$50.40$[$\ell$~Car]. This means that, in both cases, the
extrapolation to the photosphere is verified. We thus have $\Delta
V_{\mathrm{p}}^{\mathrm{g}}=p_{\mathrm{o}} b_{\mathrm{o}}$. This is
the key point that shows that the method proposed here is consistent
and can be applied to any Cepheid. To allow this very important
condition, one has to use the spectral line depth corresponding to
the minimum extension of the star ($D \equiv
D(\phi=R_{\mathrm{min}})$). If another estimator is used, for
example $D \equiv D(\phi=R_{\mathrm{max}})$, or the line depth
averaged over the entire pulsation cycle $D \equiv <D>$, then the
condition $\Delta V_{\mathrm{p}}^{\mathrm{g}}=p_{\mathrm{o}}
b_{\mathrm{o}}$ disappears. Indeed, by extrapolating the three
$p_{\mathrm{o}} \Delta RV_{\mathrm{c}}$ linear relations
corresponding to the three estimators ($D(\phi=R_{\mathrm{max}})$,
$<D>$, and $D(\phi=R_{\mathrm{min}})$, respectively) towards the
zero line depth, we find different agreements compared to the
reference values derived directly from the variation in the
photospheric pulsation velocity of the model: $-1.13$\%, $-0.70$\%,
and $0.09$\% for $\delta$~Cep and $-4.53$\%, $-3.29$\%, and $0.25$\%
for $\ell$~Car (see Fig.~\ref{Fig_depth}). The best extrapolation is
thus obtained for the `minimum' estimator. We emphasize that this
result is verified for both models ($\delta$~Cep and $\ell$~Car)
that have very different physical properties : in period ($5.4$d,
$35.5$d), radius ($43.4\,\Ro$, $180\,\Ro$), mass ($4.8\,\Mo$,
$11.5\,\Mo$), and effective temperature ($5877 K$, $5225 K$). The
`minimum' estimator of the line depth ($D$) is used for all Cepheids
in the whole paper. Considering this estimator (link to the minimum
extension of the star) does not mean that our projection factors are
related to any specific pulsation phase. It only means that we have
to use this estimator to get the right extrapolation of the
amplitude (peak-to-peak) of the pulsation velocity toward the
photosphere.
\end{enumerate}

Thus, taking the linear correlation $\Delta RV_{\mathrm{c}} = a_0 D
+ b_0$ into account, we propose the following relation to
observationally derive $f_{\mathrm{grad}}$ :

\begin{equation} \label{Eq_method}
f_{\mathrm{grad}}= \frac{b_0}{a_0D+ b_0}
\end{equation}
This quantity can be derived directly from observations. In the case
of our models and using the line depth corresponding to \ion{Fe}{I}
4896.439 \AA\,, we find $f_{\mathrm{grad}}=0.990$ [$\delta$~Cep] and
$f_{\mathrm{grad}}=0.988$ [$\ell$~Car]. These results are very
consistent with the same quantities derived directly from our
projection factor decomposition. Consequently, translating these
quantities into the projection factor $p= p_{\mathrm{o}}
f_{\mathrm{grad}} f_{\mathrm{o-g}}$, we find the reference values.
We have thus found a new way to determine $f_{\mathrm{grad}}$
directly from observations.

\subsection{Hydrodynamic modeling of $f_{o-g}$}\label{ss_fog}

The third quantity $f_{\mathrm{o-g}}$ is difficult to determine
directly from observations. It requires a precise knowledge of the
dynamical structure of the Cepheid atmosphere and line-forming
regions. In the cases of $\delta$~Cep and $\ell$~Car, the model
gives $f_{\mathrm{o-g}}=0.963$ and $0.944$, respectively.

In order to test the $P f_{\mathrm{o-g}}$ relation, we modeled the
other Cepheids of our sample (see Table
\ref{Tab_Cepheids_Parameters}). We only consider the velocity curves
corresponding to the {\it optical} and {\it gas} photospheric
layers. Consequently, no radiative transfer is calculated, and the
resulting physical parameters of these models should be considered
with caution. Nevertheless, these uncertainties are not critical for
deriving $f_{\mathrm{o-g}}$. The results are 0.966[S~Cru],
0.962[Y~Sgr], 0.955[$\beta$~Dor], 0.953[$\zeta$~Gem], 0.951[RZ~Vel],
and 0.943[RS~Pup]. We consider a 0.05 uncertainty on $f_{o-g}$.

\begin{figure}[]
\resizebox{\hsize}{!}{\includegraphics[clip=true]{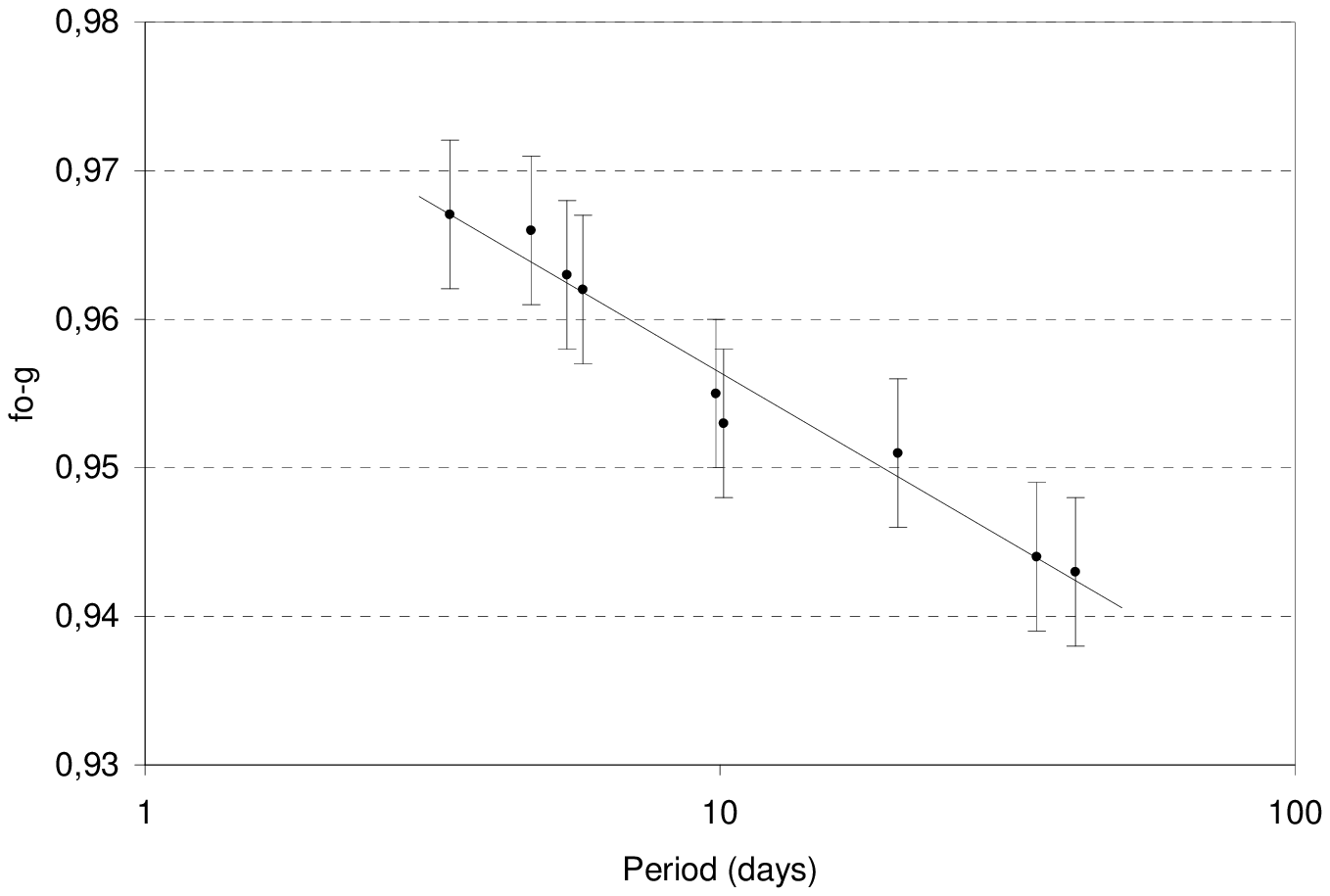}}
\caption{$f_{o-g}$ as a function of the period} \label{Fig_Pfog}
\end{figure}

Using these results (based on eight models), we obtain the following
linear relation:

\begin{equation} \label{Eq_fog}
f_{o-g}= [-0.023 \pm 0.005] \log P + [0.979 \pm 0.005]
\end{equation}
Using this relation we find $f_{o-g}=0.967$ for R~TrA. $f_{o-g}$ is
plotted as a function of the period in Fig.~\ref{Fig_Pfog}.

This linearity can be understood with the following picture. Let us
assume two atmospheres of short- and long-period Cepheids (noted S
and L) at expansion. The {\it gas} velocity is assumed to be the
same for both atmospheres. Then, due to a geometrical effect (the
radius of L is larger than the radius of S), the volume of L
increases faster than S. Consequently, the density and temperature
(in the adiabatic limit) also decrease faster in L. The lower atomic
level of this line depopulates faster for L than for S, the opacity
in the line also falls faster, the spectral line forms lower in the
atmosphere (closer to the photosphere), and finally, the {\it
optical} radius increases less. As a consequence, the velocity of
the {\it optical } layer decreases for a long-period Cepheid, while
the {\it gas} velocity is supposed to be the same. This picture can
be generalized at contraction and for the photospheric layer. Thus,
$f_{\mathrm{o-g}}=\frac{\Delta V_{\mathrm{p}}^{\mathrm{o}}}{\Delta
V_{\mathrm{p}}^{\mathrm{g}}}$ decreases with the period of the
Cepheid. Even if the $P f_{\mathrm{o-g}}$ relation seems secured, we
keep in mind that it must be studied in detail going further in our
understanding of the dynamical structure of Cepheid's atmosphere.
Important links between $f_{o-g}$, the $\gamma$-velocity, line
asymmetry, and velocity gradients should be found.

\section{Direct measurement of velocity gradients from HARPS
observations}\label{s_observations}

This section only deals with observations. For the eight stars
observed, we derive $RV_{\mathrm{c}}$ and the line depth as a
function of the pulsation phase for all spectral lines of Table
\ref{Tab_Lines}. Corresponding uncertainties are estimated based on
the signal-to-noise ratio.

\begin{figure*}[]
\resizebox{\hsize}{!}{\includegraphics[clip=true]{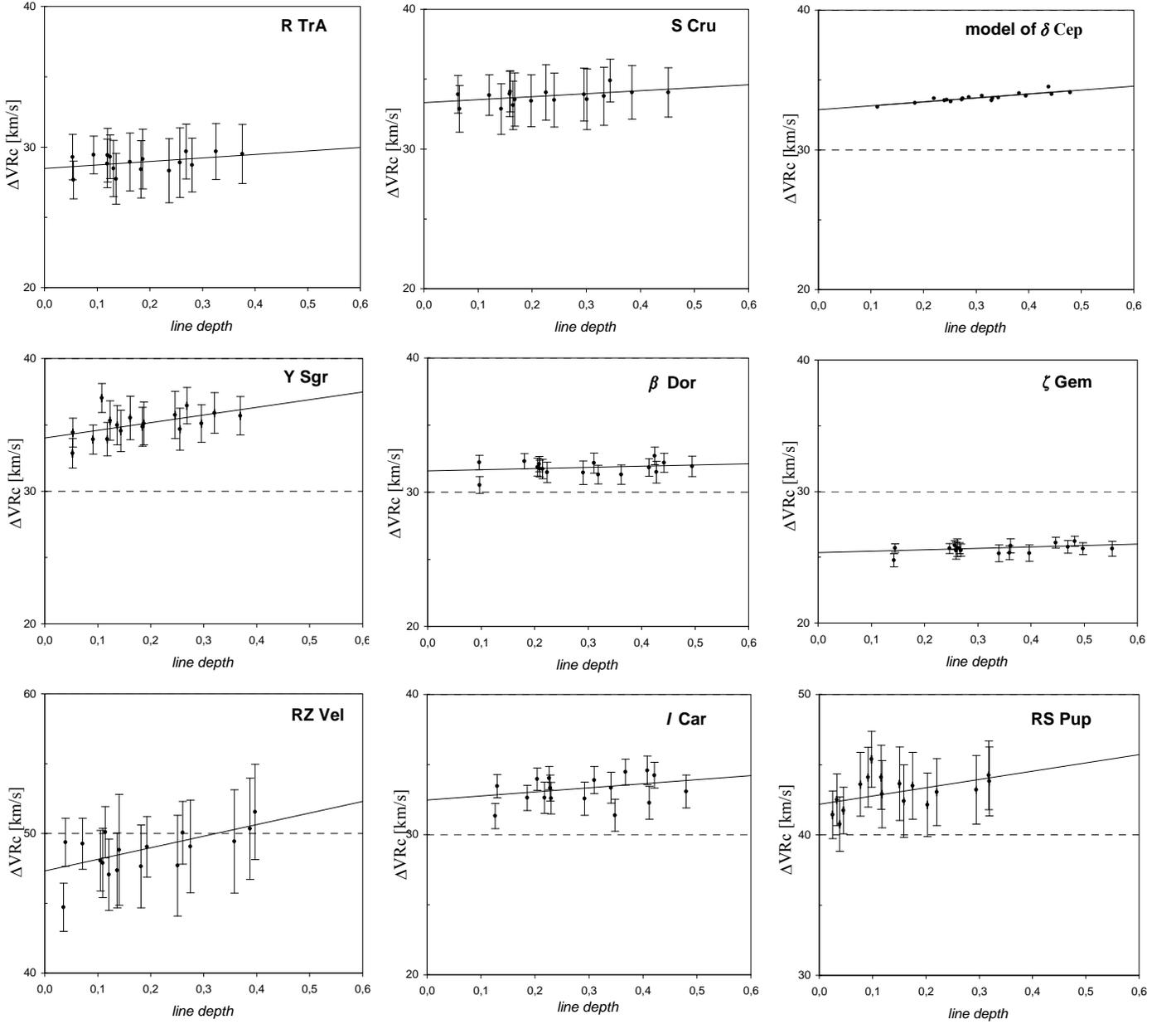}}
\caption{${\Delta RV_{\mathrm{c}}}$ as a function of the depth $D$
of the spectral line considered. Uncertainties are indicated. Stars
are presented with increasing period. The $\delta$~Cep model is also
indicated for comparison. Linear correlations are derived for all
stars (see Tab. \ref{Tab_Results}). The $f_{\mathrm{grad}}$ quantity
is derived from these relations.} \label{Fig_A_DVRc}
\end{figure*}

From interpolated curves (through periodic cubic spline functions),
we derive $\Delta RV_{\mathrm{c}}=$ {\footnotesize
MAX}$(RV_{\mathrm{c}})-$ {\footnotesize MIN}$(RV_{\mathrm{c}})$ for
all stars and lines. We also determine the pulsation phase
corresponding to the minimum extension of the stars in order to
derive $D$. $\Delta RV_{\mathrm{c}}$ is represented as a function of
line depth $D$ for all stars in Fig.~\ref{Fig_A_DVRc}. Significant
linear relations (see Table \ref{Tab_Results}) are found between
these two quantities. The amplitude of the velocity curves increases
with the line depth (or with the position of the line-forming region
in the atmosphere). This validates {\it a posteriori} the use of the
line depth as an estimator of the line-forming region. From the
coefficients provided in Table\,\ref{Tab_Results}, we are now able
to derive the $f_{\mathrm{grad}}$ quantity for each star using Eq.
\ref{Eq_method}.

\begin{table}
\begin{center}
\caption{Linear relations between the amplitude of the velocity
curves and the line depth (at minimum extension of the star),
${\Delta RV_c}= a_0  D + b_0 $ are given for all stars, together
with the $1\sigma$ uncertainty. The {\it reduced} $\chi^2$, defined
as $\chi_{\mathrm{red}}^2=\frac{\chi^2}{N-\nu}$ with N the number of
spectral lines and $\nu$ the number of degrees of freedom is also
indicated. \label{Tab_Results}}
\begin{tabular}{lccc}
\hline \hline \noalign{\smallskip}

Star    &   $a_0$    &   $b_0$    &    $\chi_{\mathrm{red}}^2$  \\
    \hline
R~TrA   &   2.50 $\pm$ 4.55    &   28.48  $\pm$   0.90    &   2   \\
S~Cru   &   2.13 $\pm$ 3.61    &   33.33  $\pm$   0.90    &   2   \\
Y~Sgr   &   5.76 $\pm$ 3.53    &   34.01  $\pm$   0.12    &   20  \\
$\beta$~Dor &   0.86  $\pm$ 1.31    &   31.59  $\pm$   0.40    &   9   \\
$\zeta$~Gem &   1.09  $\pm$   0.89    &   25.35 $\pm$   0.31    &   7   \\
RZ~Vel  &   8.32   $\pm$   5.95    &   47.31   $\pm$   1.02    &   6   \\
$\ell$~Car  &   2.89    $\pm$   2.26    &   32.48   $\pm$   0.67    &   15  \\
RS~Pup  &   5.89    $\pm$   5.58    &   42.19   $\pm$   0.88    &   4   \\
\hline \noalign{\smallskip}
\end{tabular}
\end{center}
\end{table}

However, if one wants to compare the velocity gradient within the
atmosphere of different Cepheids, the question of methodology
arises. Two strategies are possible. One can consider (1)~the same
line depth for all stars or (2)~the same spectral line. Because  a
given spectral line does not have the same depth for all stars, the
choice is important. We first use strategy (1) to determine which
spectral lines are most fitting for the IBW method, and then
consider a specific spectral line to compare the velocity gradient
in the Cepheids' atmosphere.

\begin{figure}[]
\resizebox{\hsize}{!}{\includegraphics[clip=true]{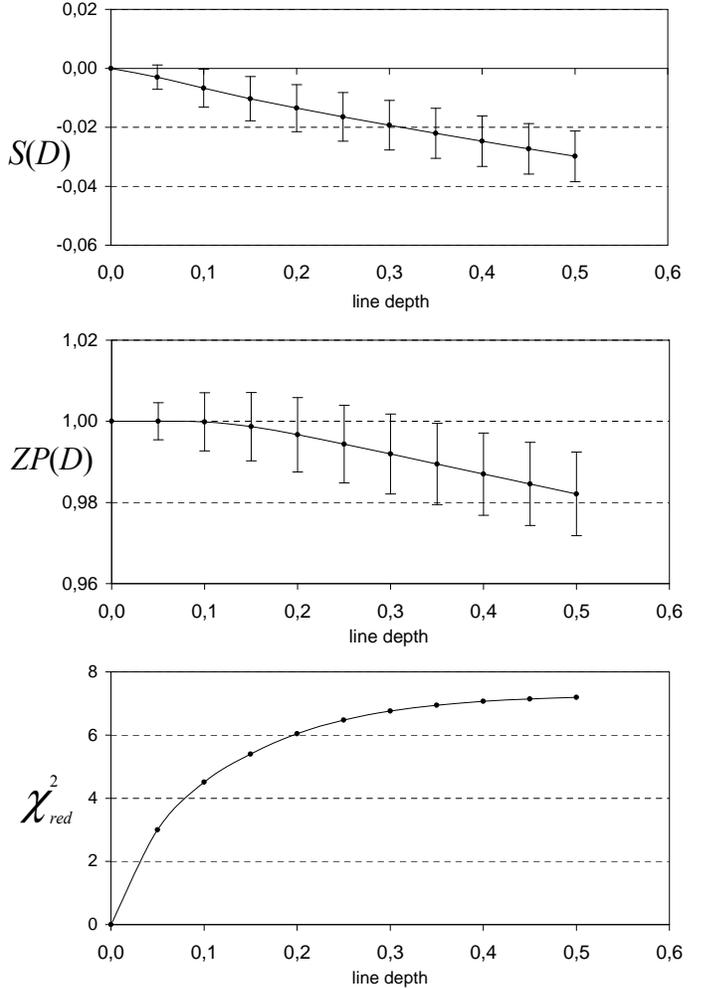}}
\caption{Slope and zero-point of the Period-$f_{\rm grad}$ relation
($f_{\rm grad}=S(D) \log P + ZP(D)$) as a function of the line
depth. The reduced $\chi^2$ is also indicated.} \label{Fig_Strategy}
\end{figure}

First, we test the existence of a linear relation between
$f_{\mathrm{grad}}=\frac{b_0}{a_0D+ b_0}$ and the logarithm of the
period ($\log P$) for eleven different line depths from 0 to 0.5 in
increments of 0.05. By a minimization process, we obtain the
relations: $f_{\mathrm{grad}}=S(D) \log P + ZP(D)$, where the slope
($S$) and the zero-point ($ZP$) are related to the line depth
considered. The variables $S$, $ZP$, and the reduced $\chi^2$ are
represented as a function of the line depth in
Fig.~\ref{Fig_Strategy}. The physical interpretation of these curves
is as follows.

First, for a line depth of zero, the velocity gradient is also zero
for all stars (or $f_{\mathrm{grad}}=1$). Thus, $S=0$ , $ZP=1$, and
there is no uncertainty on $f_{\mathrm{grad}}$. The reduced $\chi^2$
is 0 since the linear relation is perfect.

Then, for low depths, the $f_{\mathrm{grad}}$ estimator is quite
different for each star, but the effect is low and the
$f_{\mathrm{grad}}=S(D) \log P + ZP(D)$ relation is still not very
sensitive to the velocity gradient. Moreover, the uncertainty on
$f_{\mathrm{grad}}$ is low, and the same is true for $S$ and $ZP$.
The reduced $\chi^2$ is good, as the linearity is well-conserved.

However, for large depths, the $f_{\mathrm{grad}}$ estimator becomes
more and more sensitive (but also more uncertain) to the velocity
gradient. The dispersion between the different stars is amplified.
As a consequence, the linearity is verified at a lower level: the
reduced $\chi^2$ gets larger.

From this picture, our objective is to find the best spectral lines
to use in determining $f_{\mathrm{grad}}$ (and then the projection
factor) in the context of the IBW method. Spectral lines with depths
lower than 0.1 seem to be the best choice. Such spectral lines are
indeed less sensitive to the velocity gradient. It is obvious that a
spectral line that forms close to the photosphere implies small
differences (in velocity) between the line-forming region and the
photosphere. In this case, $f_{\mathrm{grad}}$ is close to $1$ and
the corresponding uncertainty is low. Indeed, from
Fig.~\ref{Fig_Strategy}, we clearly see that the uncertainties on
$S$ and $ZP$ are decreasing for low depths. From
Table\,\ref{Tab_Lines}, we find that the \ion{Fe}{I} 4896.439 \AA\
spectral line is the best. It has the lowest depth, averaged over
all pulsation phases for all stars ($D=0.08$).

\begin{figure}[]
\resizebox{\hsize}{!}{\includegraphics[clip=true]{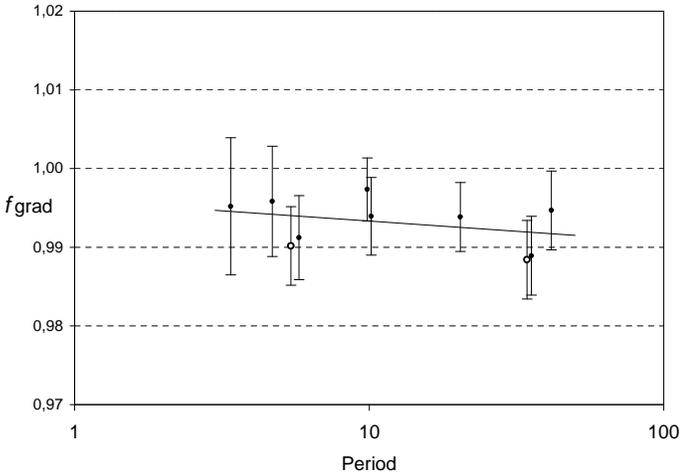}}
\caption{$f_{\mathrm{grad}}$ in the case of the \ion{Fe}{I} 4896.439
spectral line is represented as a function of the period. Open
circles corresponds to the hydrodynamic models. The slope and the
zero-point of the relation depend on the line depth (see
Fig.~\ref{Fig_Strategy}).} \label{Fig_fgrad}
\end{figure}

In Fig.~\ref{Fig_fgrad} we present $f_{\mathrm{grad}}$ as a function
of the logarithm of the period for the \ion{Fe}{I} 4896.439 \AA\
spectral line. We obtain a linear relation between the velocity
gradient and the logarithm of the period. Results are given in Table
\ref{Tab_Results_fgrad} for observations only (O), and for
observations + the $\delta$~Cep and $\ell$~Car hydrodynamical models
(O+C). Consequently, we can conclude that the velocity gradient is
larger in long-period Cepheids than in short-period Cepheids.
Moreover, if we compare this observational result to the
hydrodynamical models of $\delta$~Cep and $\ell$~Car, we find very
good agreement (see Fig.~\ref{Fig_fgrad}). The two models presented
are thus extremely good.

\begin{table}
\begin{center}
\caption{Linear relation between $f_{\mathrm{grad}}$ and the
logarithm of the period : $f_{\mathrm{grad}} = S \log P + ZP$ for
the \ion{Fe}{I} 4896.439 \AA\ spectral line.
\label{Tab_Results_fgrad}}
\begin{tabular}{lccccc}
\hline \hline \noalign{\smallskip}

                       &  S      & ZP     &
                       $\chi_{red}^2$ \\
\hline
 O      & -0.003 $\pm$ 0.005   & 0.997 $\pm$ 0.006  & 2 \\
 O + C  & -0.003 $\pm$ 0.004 & 0.996 $\pm$ 0.005 &
 3 \\

\hline \noalign{\smallskip}
\end{tabular}
\end{center}
\end{table}

\section{A period-projection factor relation} \label{s_Pp}

\begin{table*}
\begin{center}
\caption[]{Derived projections factors for all stars computed from the
decomposition presented in Eq. \ref{Eq_pf_decomposition}.
\label{Tab_presults}}
\begin{tabular}{lcccccccccc}
\hline \hline \noalign{\smallskip}

Name     &  HD     &  $P$~{\tiny (b)}   &         $p_{\mathrm{o}}$~{\tiny (c)}             &       $f_{\rm grad}$~{\tiny (d)}       & $f_{o-g}$~{\tiny (e)}      & $p$~{\tiny (f)} \\
         &         &  [days]            &                           &                        &                                             \\
\hline
R TrA       & 135592  &   $3.38925$     &  $1.396_{\pm 0.010}$         &   $0.995_{\pm 0.009}$     &   $0.967_{\pm 0.005}$     &     $1.34_{\pm 0.03}$ \\
S Cru       & 112044  &   $4.68976$     &  $1.392_{\pm 0.010}$         &   $0.996_{\pm 0.007}$      &   $0.966_{\pm 0.005}$     &  $1.34_{\pm 0.03}$   \\
Y Sgr       & 168608  &   $5.77338$     &  $1.387_{\pm 0.010}$         &   $0.991_{\pm 0.005}$  &   $0.962_{\pm 0.005}$      & $1.32_{\pm 0.02}$  \\
$\beta$ Dor & 37350   &   $9.84262$     &  $1.380_{\pm 0.010}$         &   $0.997_{\pm 0.004}$    &   $0.955_{\pm 0.005}$    & $1.31_{\pm 0.02}$   \\
$\zeta$ Gem & 52973   &   $10.14960$    &  $1.380_{\pm 0.010}$         &   $0.994_{\pm 0.005}$       & $0.953_{\pm 0.005}$      & $1.31_{\pm 0.02}$   \\
RZ Vel      & 73502   &   $20.40020$    &  $1.375_{\pm 0.010}$         &   $0.994_{\pm 0.004}$       & $0.951_{\pm 0.005}$      & $1.30_{\pm 0.02}$  \\
$\ell$~Car  & 84810   &   $35.55134$   &  $1.366_{\pm 0.010}$         &   $0.989_{\pm 0.005}$       & $0.944_{\pm 0.005}$   & $1.27_{\pm 0.02}$  \\
RS Pup      & 68860   &   $41.51500$    &  $1.360_{\pm 0.010}$         &   $0.995_{\pm 0.005}$       & $0.943_{\pm 0.005}$     & $1.28_{\pm 0.02}$ \\
\hline
$\delta$~Cep ~{\tiny (a)}& 213306   &   $5.419$  &   $1.390_{\pm 0.010}$  & $0.990_{\pm 0.005}  $     &   $0.963_{\pm 0.005}$      & $1.33_{\pm 0.02}$ \\
$\ell$~Car ~{\tiny (a)}  & 84810   &    $35.60$  &    $1.366_{\pm 0.010}$ & $0.988_{\pm 0.005} $      &   $0.944_{\pm 0.005}$     & $1.27_{\pm 0.02}$  \\

\hline \noalign{\smallskip}
\end{tabular}
\end{center}
\begin{list}{}{}

\item[$^{\mathrm{a}}$] $\delta$~Cep and $\ell$~Car are hydrodynamical models

\item[$^{\mathrm{b}}$] The corresponding Julian dates
($T_{\mathrm{o}}$) can be found in Paper II.

\item[$^{\mathrm{c}}$] $p_{\mathrm{o}}$ is derived from the linear limb-darkening laws of Claret et al. (2000)
based on the static models of Kurucz (1992). We then apply a slight
correction based on the $\delta$~Cep and $\ell$~Car hydrodynamical
models: $p_0$[hydro]$=p_0$[geo]$-(0.0174 \log P -0.0022)$ to take
the dynamical structure of the Cepheid's atmosphere into account.

\item[$^{\mathrm{d}}$] $f_{\rm grad}$ is derived directly from observations
using Eq \ref{Eq_method}. It is important to notice that the results
indicated here correspond to the \ion{Fe}{I} 4896.439 \AA\ line. In
the case of a modeled star, it is derived directly from the
hydrodynamical model (see Sect.~\ref{sss_method}).

\item[$^{\mathrm{e}}$] $f_{o-g}$ is derived directly from the hydrodynamical
models (see Sect. \ref{ss_fog}).

\item[$^{\mathrm{f}}$] $p$-factors defined by
$p=p_{\mathrm{o}}f_{\mathrm{grad}}f_{\mathrm{o-g}}$.
$p_{\mathrm{o}}$ and $f_{\mathrm{o-g}}$ are derived from geometrical
and hydrodynamical models respectively. $f_{\mathrm{grad}}$ is
derived from observations.
\end{list}

\end{table*}

We now determine the projection factors of Cepheids considering the
\ion{Fe}{I} 4896.439~\AA\ spectral line. The
$p_{\mathrm{o}}$-factors are not determined directly using the
continuum intensity distribution of the hydrodynamical models. As
explained in Sect. \ref{ss_fog}, these models (except for
$\delta$~Cep and $\ell$~Car) are only used to derive
$f_{\mathrm{o-g}}$; no radiative transfer in the atmosphere is
computed, and the observed spectral lines profiles (and radial
velocity curves) are not used to constrain the models. Consequently,
we are not certain about the dynamical structure of the modeled
Cepheids' atmosphere and, in particular, about $f_{\mathrm{grad}}$.
And if $f_{\mathrm{grad}}$ is wrong, the limb darkening within the
line (and thus the $p_{\mathrm{o}}$-factor) could be affected.
Consequently, we consider a conservative approach when deriving
$p_o$[geo] from the hydrodynamic parameters of
Table~\ref{Tab_Cepheids_Parameters}. We then apply a slight
correction: $p_0$[hydro]$=p_0$[geo]$-(0.0174 \log P -0.0022)$ using
the results based on $\delta$~Cep and $\ell$~Car hydrodynamic
models, which were studied in detail (Sect.~\ref{ss_modeling}). The
resulting linear relation is

\begin{equation}
p_{\mathrm{o}} = [-0.031 \pm 0.008] \log P + [1.413 \pm 0.009].
\end{equation}

\begin{figure}[]
\resizebox{\hsize}{!}{\includegraphics[clip=true]{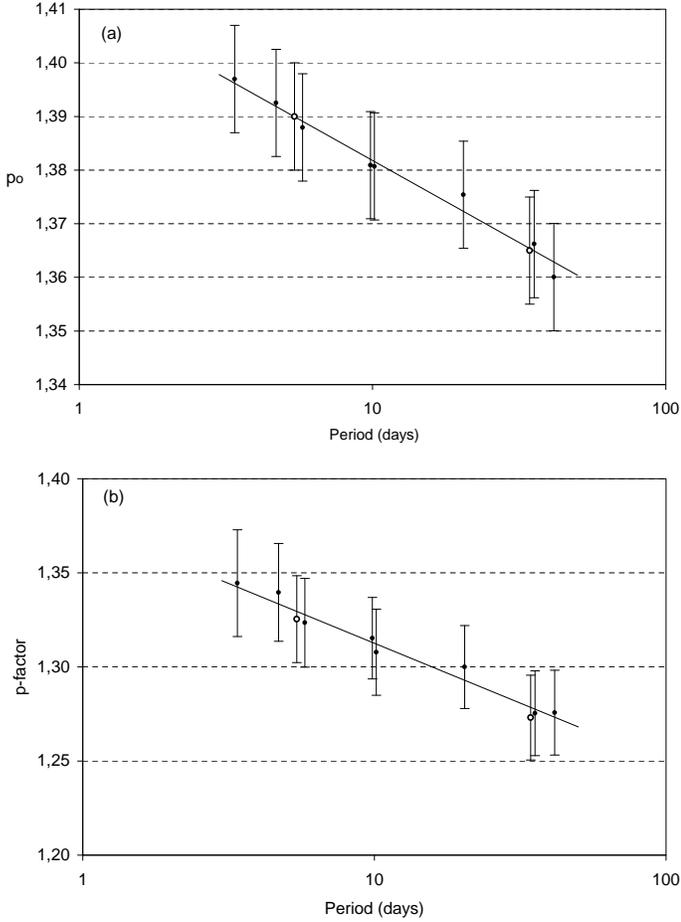}}
\caption{ The projection factor ($p=p_{\mathrm{o}}f_{\mathrm{
grad}}f_{\mathrm{o-g}}$) as a function of the logarithm of the
period (diag. b), together with $p_{\mathrm{o}}$ (diag. a). Black
points and open circles correspond to observations and models.}
\label{Fig_Pp}
\end{figure}

The $f_{\mathrm{o-g}}$ quantities have been deduced directly from
the hydrodynamical models (see Sect. \ref{ss_fog}), and
$f_{\mathrm{grad}}$ was determined from observations in the previous
section. From $p_{\mathrm{o}}$, $f_{\mathrm{o-g}}$, and
$f_{\mathrm{grad}}$, we can now determine consistent projection
factors for all stars. A summary of the results is given in Table
\ref{Tab_presults} and illustrated in Fig.~\ref{Fig_Pp}.

By combining all quantities ($p_{\mathrm{o}}$, $f_{\mathrm{grad}}$
and $f_{\mathrm{o-g}}$), we are able to derive a \emph{Pp} relation
for the first time (see Fig.~\ref{Fig_Pp}-b). This result is a
combination of observed and theoretical considerations. The
resulting linear law (including observed and modeled stars) is:

\begin{equation} \label{Eq_P_p_O_C}
p = [-0.064 \pm 0.020] \log P + [1.376 \pm 0.023].
\end{equation} This relation holds for the \ion{Fe}{I}~4896.439 \AA\ spectral line,
which presents the lowest line depth in our sample.

From this relation, two facts must be pointed out. First, the
$Pp_{\mathrm{o}}$ relation is based on the general physical
properties of Cepheids (effective temperature, surface gravity),
while $P f_{\mathrm{grad}}$ is derived directly from observations
and states that the velocity gradient within the atmosphere is
larger in long-period Cepheids. The linearity of the $P
f_{\mathrm{o-g}}$ relation is qualitatively understood (Sect.
\ref{sss_method}) and has been quantitatively verified based on
eight models. Second, results concerning the $\delta$~Cep and
$\ell$~Car hydrodynamic models are highly secured: (1) the
projection factor of $\delta$~Cep was confirmed observationally by
M\'erand et al. (2005); (2) the velocity gradient in the atmosphere
of these stars is confirmed by HARPS observations (see
Sect.~\ref{s_observations}); (3) the $p_{\mathrm{o}}$ estimations
are coherent at a 0.025 level with the geometrical models; and (4)
from points 1, 2, and 3, we can reasonably feel secure in the
estimations of $f_{\mathrm{o-g}}$ for $\delta$~Cep, $\ell$~Car, and
for all stars.

\section{Discussion}

The derived \emph{Pp} relation will be useful in the context of the
IBW and SB methods. For example, if we compare Eq. \ref{Eq_P_p_O_C}
with the usual value widely used in the community $p=1.36$ (Burki et
al. 1982), we obtain a correction for the projection factor
depending on the period. It is then possible to translate it into a
bias on distances and absolute magnitudes. By this process, we
obtain the relation:

\begin{equation} \label{Eq_P_po}
\Delta M_V = 0.10 \log P - 0.03
\end{equation} where $\Delta M_V$ is the correction to consider on the \emph{PL}
relation. We thus conclude that one can make an errors of $0.10$ and
$0.03$ on the slope and zero-point of the \emph{PL} relation,
respectively, if $p=1.36$ is used for all stars instead of the
\emph{Pp} relation. This correction is, however, only indicative
because it is indeed restricted to our definition of the projection
factor (Eq. \ref{Eq_pf}) and to the \ion{Fe}{I} 4896.439 \AA\
spectral line.

It is now possible to refine the IBW and SB methods. First, we
suggest using the $RV_{\mathrm{c}}$ radial velocity to avoid bias
related to the rotation velocity of the star (even if Cepheids are
supposed to be slow rotators) and the width of the spectral line.
One then has to determine the $RV_{\mathrm{c}}$ curve and force the
average to be zero in order to avoid $\gamma$-velocity effects. Due
to our careful definition of $p$ (Eq.~\ref{Eq_pf}), the projection
factors proposed in this paper are indeed independent of the
${\gamma}$-velocity. The spectral line considered must have a depth
lower than 0.1 and should be the same for all considered Cepheids.
The low depth of the spectral line is required to diminish the
impact of the velocity gradient. If the \ion{Fe}{I} 4896.439 \AA\ is
used, one can use Eq.~\ref{Eq_P_p_O_C} directly to determine the
dynamic projection factors of Cepheids. If not, we propose the
following method. Given the line depth of the spectral line
considered for each Cepheid, it is possible to determine the
$f_{\mathrm{grad}}$ from Table\,\ref{Tab_Results} and Eq.
\ref{Eq_method}. If the Cepheid being studied is not in our sample,
Fig.~\ref{Fig_Strategy} can be used. Then $p_{\mathrm{o}}$ can be
determined using a geometrical model. However, the consistency (at a
level lower than $0.025$ on $p$) between interferometric
observations, geometrical, and hydrodynamical models should be
studied in detail in the future. For $f_{\mathrm{o-g}}$, one can use
Eq. \ref{Eq_fog}, even if this relation has to be confirmed
observationally in the future. For this purpose, the development of
theory and hydrodynamical models is required. Finally, the
projection factor of Cepheids, following our decomposition, is
$p=p_{\mathrm{o}}\,f_{\mathrm{grad}}\,f_{\mathrm{o-g}}$. This
procedure should be applied to avoid bias in the calibration of the
\emph{PL} relation.

However, we know that the masking cross-correlation method is widely
used to increase the signal-to-noise ratio on radial velocity
measurements. In that case however, one cannot exclude the impact of
the rotation, the spectral lines' width, and $\gamma$-velocities
effects. Nevertheless, we can still provide a \emph{Pp} relation
that is more appropriate considering an average line depth of
$D=0.25$. We find $p= [-0.075\pm0.031] \log P + [1.366\pm0.036]$.

Another important point is that we provide visible projection
factors that should be used with visible spectroscopic observations.
If one used infrared spectroscopic observations to derive the
pulsation velocity, one should use specific infrared projection
factors. Indeed, in the infrared, the limb darkening is supposed to
be lower and the corresponding $p_{\mathrm{o}}$-factors higher
(certainly about 4\%). But, spectral lines also form higher in the
atmosphere (i.e. in the upper part of the atmosphere), which
supposes a lower $f_{\mathrm{grad}}$. More studies have to be
carried out to derive an infrared \emph{Pp} relation.

\section{Conclusion}

In the application of the IBW method, the projection factor is a key
quantity. Up to now, the period-dependency of the projection factor
has never been studied in detail. Here, we have presented a new
spectroscopic method for directly measuring the velocity gradient in
the Cepheids' atmosphere. This method has been successfully
validated by the hydrodynamical models of $\delta$~Cep and
$\ell$~Car. We find a physical relation between the period of the
star and its dynamical atmospheric structure.

The models also show that the {\it optical} layers (observed by
continuum interferometer) and the {\it gas} layers have to be
distinguished in the interferometric definition of the projection
factor. However, this quantity is still very difficult to determine
directly from observations.

Combining the results obtained directly from observations and our
knowledge of the dynamical structure of the $\delta$~Cep and
$\ell$~Car atmosphere, we have been able to derive a very consistent
\emph{Pp} relation for the \ion{Fe}{I} 4896.439 \AA\ spectral line:

\begin{equation} \label{Eq_P_po}
p = [-0.064 \pm 0.020] \log P + [1.376 \pm 0.023].
\end{equation}

We emphasize that, if a constant projection factor is used to
constrain the \emph{PL} relation, an error of $0.10$ and $0.03$
magnitudes can be done, respectively, on the slope and zero-point of
the \emph{PL} relation. This can even be much more if the wrong
definition of the radial velocity is used or if one does not
consider $\gamma$-velocity effects. We have thus presented (see
discussion) a careful methodology to be applied in the context of
the IBW and SB methods.

\begin{acknowledgements}
Based on observations collected at La Silla observatory, Chile, in
the framework of European Southern Observatory's programs 072.D-0419
and 073.D-0136. This research made use of the SIMBAD and VIZIER
databases at the CDS, Strasbourg (France). We thank C. Catala for
useful discussions of the line-forming region estimator, P. Kervella
for having provided the HARPS data and M. Fekety as well as Joli
Adams for their careful English correction of the paper. N. Nardetto
acknowledges the Max Planck Institut for Radioastronomy for
financial support.

\end{acknowledgements}


\end{document}